\documentclass[aps,superscriptaddress,prx,11pt,longbibliography,tightenlines]{revtex4}
\usepackage[T1]{fontenc}
\usepackage{newtxtext}
\usepackage{amsmath}

\usepackage{mathtools}
\usepackage{amsfonts}
\usepackage{bm}
\usepackage{booktabs}
\usepackage{makecell}
\usepackage{color}
\usepackage{xcolor}
\definecolor{auburn}{rgb}{0.43, 0.21, 0.1}
\definecolor{blue}{rgb}{0.2, 0.3, 0.85}
\definecolor{red}{rgb}{0.95, 0.1, 0.15}
\usepackage[colorlinks,bookmarks=false,citecolor=blue,linkcolor=auburn,urlcolor=blue]{hyperref}
\def\be{\begin{equation}}
\def\ee{\end{equation}}
\def\bea{\begin{eqnarray}}
\def\eea{\end{eqnarray}}

\begin{document}
\title{Entanglement entropy and localization in disordered quantum chains}
\author{Nicolas Laflorencie}
\affiliation{Laboratoire de Physique Th\'eorique, Universit\'e de Toulouse, CNRS, UPS, France}
\affiliation{Donostia International Physics Center, Paseo Manuel de Lardizabal 4, E-20018 San Sebasti\'an, Spain}
\begin{abstract}
This chapter addresses the question of quantum entanglement in disordered chains, focusing on the von-Neumann and R\'enyi entropies for three important classes of random systems: Anderson localized, infinite randomness criticality, and many-body localization (MBL). We review previous works, and also present new results for the entanglement entropy of random spin chains at low and high energy.
\end{abstract}
\maketitle
\vskip -0.75cm
\tableofcontents
\section{Introduction}
\label{sec:introduction}
\subsection{Generalities}
Random impurities, disorder, and quantum fluctuations have the common tendency to conspire, destroy classical order, and drive physical systems towards new states of matter. Whether
intrinsically present, chemically controlled via doping materials, or
explicitly introduced via a random potential (as in ultra-cold atomic setups) of for instance by varying 2D film thickness,
randomness can lead to dramatic changes in many properties of condensed
matter systems, as exemplified by Anderson localization phenomena~\cite{anderson_absence_1958,evers_anderson_2008}, the Kondo effect~\cite{kondo_resistance_1964,hewson_kondo_1993}, or spin-glass physics~\cite{binder_spin_1986}. In such a context, the introduction of quantum entanglement witnesses provides new tools to improve our  understanding of quantum disordered systems. Among the numerous entanglement estimates, one of the simplest is the so-called von-Neumann entropy, that will be described in this chapter for various one-dimensional disordered localized states of matter.
\subsection{Random spin chain models}
\subsubsection{Disordered XXZ Hamiltonians}
\paragraph*{(i) Models---} Several spin systems will be discussed along this chapter. The first (prototypical) example is the U(1) symmetric disordered spin-1/2 XXZ model
\be
{\cal{H}}_s=\sum_{i,j}J_{ij}\left(S_{i}^xS_{j}^x+S_{i}^yS_{j}^y+\Delta S_{i}^zS_{j}^z\right)+\sum_i h_iS_i^z,
\label{eq:Hs}
\ee
where the total magnetization is conserved $\left[{\cal{H}},\sum_iS_i^z\right]=0$. This Hamiltonian is quite generic as it can also describe bosonic or  fermionic systems. Indeed, using the Matsubara-Matsuda mapping~\cite{matsubara_lattice_1956} $b_i^\dagger=S_i^+$,  $b_i^{\vphantom{\dagger}}=S_i^-$, and $n_i=S_i^z+1/2$, the above spin problem Eq.~\eqref{eq:Hs} equally describes hard-core bosons 
\be
{\cal{H}}_b=\sum_{i,j}\frac{J_{ij}}{2}\left(b_{i}^{\dagger}b_{j}^{\vphantom\dagger}+b_{j}^{\dagger}b_{i}^{\vphantom\dagger}+2\Delta n_{i}n_{j}\right)+\sum_i h_in_i +{\rm{constant}}.
\ee
A fermionic version can also be obtained from the Jordan-Wigner transformation~\cite{jordan_uber_1928} which maps hard-core bosons onto spinless fermions through:
\be
c_\ell= \exp\left[i\pi \sum_{j=1}^{\ell-1}b^\dagger_j b_j^{\vphantom{\dagger}}\right]b_\ell^{\vphantom{\dagger}}\quad {\rm and}\quad c^\dagger_\ell = b^\dagger_\ell\exp\left[-i\pi \sum_{j=1}^{\ell-1}b^\dagger_j b_j^{\vphantom{\dagger}}\right].
\ee
The Jordan-Wigner string, although making the transformation non-local, ensures that $c^{\vphantom{\dagger}}_\ell$ and $c^\dagger_\ell$ satisfy anticommutation relations and are indeed fermionic operators. In one dimension, if hopping terms are restricted to nearest-neighbor,  the original XXZ spin model Eq.~\eqref{eq:Hs} takes the simple spin-less fermion form
\be
{\cal{H}}_f=\sum_{i}\frac{J_{i}}{2}\left(c_{i}^{\dagger}c_{i+1}^{\vphantom\dagger}+c_{i+1}^{\dagger}c_{i}^{\vphantom\dagger}+2\Delta n_{i}n_{i+1}\right)+h_in_i.
\ee

\begin{figure}[t!]
\includegraphics[angle=0,clip=true,width=\columnwidth]{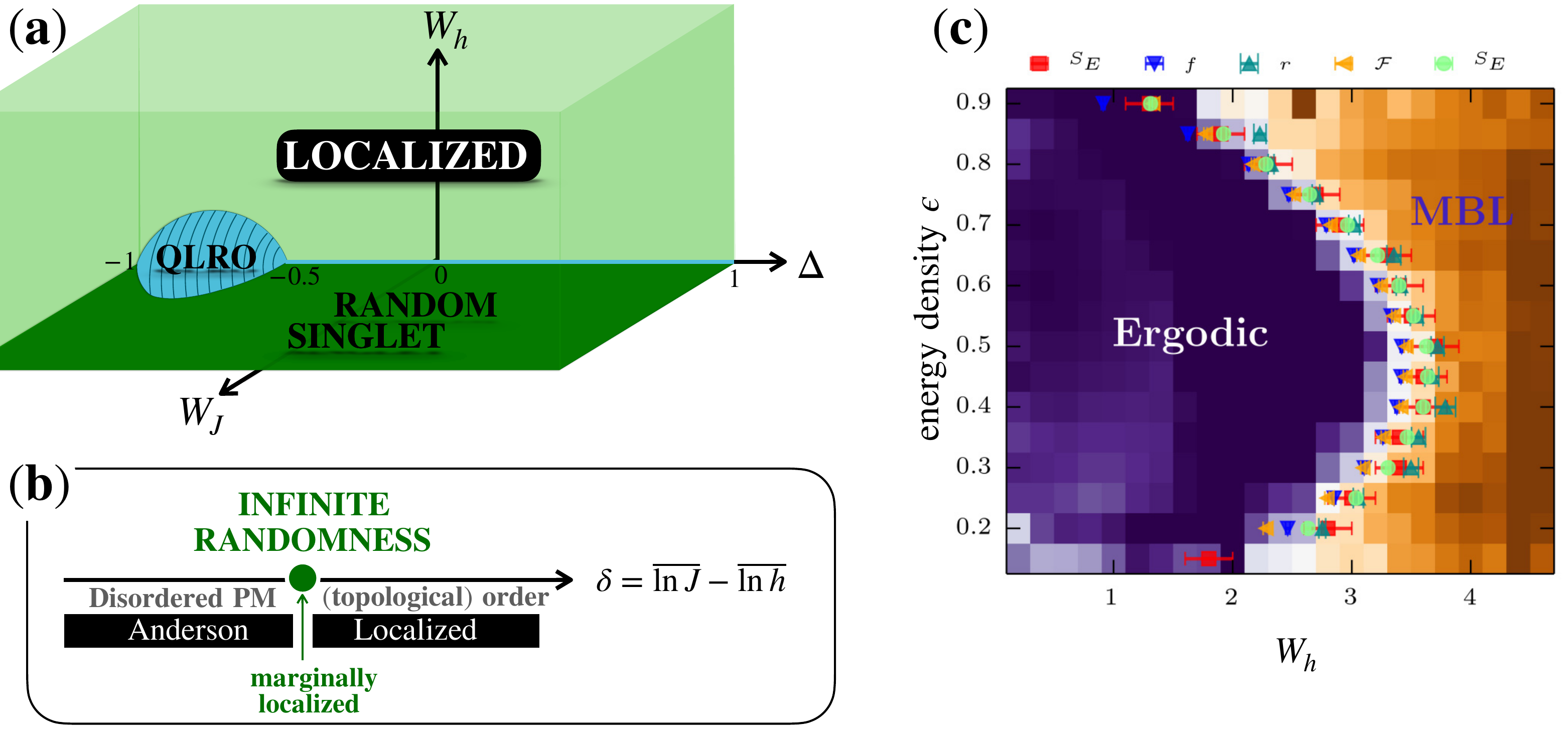}
\caption{Schematic phase diagrams for the three different systems considered. (a) Ground-state phase diagram of the disordered XXZ chain model Eq.~\eqref{eq:Hs}. $\Delta\in [-1,1]$ is the interaction parameter, and $W_J$, $W_h$ are the disorder strengths for couplings and fields (se text). Three phases are expected. In the attractive regime $\Delta\in [-1,-1/2]$, a small pocket is robust against weak disorder, showing quasi-long-range-order (QLRO). In the absence of random field ($W_h=0$) random bonds induce a random singlet phase. In the largest part of the diagram, a localized phase is expected. (b) The random one-dimensional TFIM Eq.~\eqref{eq:TFI} displays two localized phases (disordered PM and topological ordered) surrounding an infinite randomness fixed point (IRFP) at $\delta={\overline{\ln J}}-{\overline{\ln h}}=0$. (c) Energy-resoved MBL diagram for the random-field Heisenberg chain, the standard model for 1D MBL (this panel is adapted from Luitz {\it{et al.}}~\cite{luitz_many-body_2015}).}
\label{fig:gs_phase_diag}
\end{figure}

\paragraph*{(ii) Ground-state phase diagram in the presence of disorder---} Building on field-theory and renormalization-group (RG) results~\cite{giamarchi_anderson_1988,doty_effects_1992,fisher_random_1994}, as well as  numerical investigations~\cite{bouzerar_impurity_1994,schmitteckert_anderson_1998,doggen_weak-_2017,lin_many-body_2018}, the global zero-temperature phase diagram of the above disordered XXZ chain is depicted in Fig.~\ref{fig:gs_phase_diag} (a) with 3 parameters. $\Delta$ is the (non-random)  interaction strength, repulsive ($\Delta>0$) or attractive ($\Delta<0$), $\Delta=0$ being the free-fermion point ; $W_J$ controls the randomness in the antiferromagnetic exchanges $J_i>0$, which can be drawn from a power-law $P(J)\sim J^{-1+1/W_J}$ (while the precise form of the distribution is irrelevant) ; $W_h$ is the disorder strength of the random fields $h_i$, often chosen to be a uniform box $P(h)={\rm{Box}}[-W_h\,,W_{h}]$, but again its precise form is not relevant. In Fig.~\ref{fig:gs_phase_diag} (a) one sees three main regimes:
\begin{enumerate}
\item[(1)]{In the absence of randomness, and inside a small pocket (blue region), the quasi-long-range-order (QLRO) is stable, with Luttinger-liquid-like critical properties~\cite{giamarchi_quantum_2003}, such as power-law decaying pairwise correlations at long distance.}
\item[(2)]{At zero random-field ($W_h=0$), random antiferromagnetic couplings can drive the ground-state to the random singlet phase (RSP)~\cite{fisher_random_1994}: a critical glass phase controlled by an infinite randomness fixed point (IRFP)~\cite{fisher_random_1992}, having  power-law (stretched exponential) average (typical) correlations.}
\item[(3)]{IRFP and RSP are destabilized by non-zero (random) fields, driving the systems to a localized ground-state, also known as the Bose glass state~\cite{fisher_boson_1989}. This localized regime is directly connected to the non-interacting limit.}
\end{enumerate}

\subsubsection{Random transverse field Ising chains}
Another class of disordered spin chain models is given by the famous transverse-field Ising model (TFIM)
\be
{\cal{H}}_{\rm TFI}=\sum_{i}J_i\sigma^x_i \sigma^x_{i+1}+h_i \sigma^z_i,
\label{eq:TFI}
\ee
which can also be recasted into a free-fermion model 
\be
{\cal{H}}_{\rm TFI}=\sum_{i=1}^{L}\Bigl[J_i\left(c_i^\dagger c_{i+1}^\dagger+c_i^\dagger c_{i+1}^{\vphantom{\dagger}}-c_i^{\vphantom{\dagger}} c_{i+1}^{{\dagger}}-c_i c_{i+1}\right)+h_i\left(1-2c_i^\dagger c_i^{\vphantom{\dagger}}\right)\Bigr].
\label{eq:Hff}
\ee
This system is equivalent to the celebrated Kitaev chain~\cite{kitaev_unpaired_2001}, but here with equal pairing and hopping terms, and in the presence of disorder. Despite the great tour de force achieved by Kitaev who showed the non-trivial topological properties of the TFIM Eq.~\eqref{eq:TFI} (with edge Majorana zero modes, also discussed by Fendley~\cite{fendley_parafermionic_2012}), many of the properties of Eq.~\eqref{eq:Hff} were studied several decades before (in the disorder-free case) by Lieb, Schultz, Mattis~\cite{lieb_two_1961} and Pfeuty~\cite{pfeuty_one-dimensional_1970}. 

The random case, also discussed for a long time~\cite{mccoy_theory_1968,mccoy_theory_1969,fisher_random_1992,fisher_critical_1995} has been deeply understood by D. S. Fisher~\cite{fisher_random_1992,fisher_critical_1995} who solved the strong disorder renormalization group (SDRG) method for the critical point of the random TFIM at $\delta={\overline{\ln J}}-{\overline{\ln h}}=0$, which also exhibits an IRFP. This (non-interacting) quantum glass displays marginal localization for  single-particle fermionic orbitals~\cite{nandkishore_marginal_2014}, while a genuine Anderson localization is observed for $\delta\neq0$, with the following physical phases: a disordered paramagnet (PM) when $\delta<0$ and a topological ordered magnet if $\delta>0$. Physical properties of the 1D random TFIM have been studied numerically using free-fermion diagonalization techniques~\cite{young_numerical_1996,igloi_random_1998,fisher_distributions_1998} but most of these studies have focused on zero-temperature properties. Below we will address entanglement for low and (very) high energy states.

\subsubsection{Many-body localization}
\label{subsec:mbl}
Here we briefly discuss the main properties of many-body localization (MBL) physics, while referring the interested reader to recent reviews on this broadly discussed topic~\cite{nandkishore_many-body_2015,abanin_recent_2017,alet_many-body_2018,abanin_many-body_2019}. 
The excitation spectrum of disordered quantum interacting systems has been a fascinating subject for more than two decades now~\cite{jacquod_emergence_1997,gornyi_interacting_2005,basko_metalinsulator_2006,znidaric_many-body_2008,pal_many-body_2010,bardarson_unbounded_2012,imbrie_many-body_2016}. While the very first analytical studies focused on the effect of weak interactions~\cite{gornyi_interacting_2005,basko_metalinsulator_2006}, the majority of the subsequent numerical studies then addressed strongly interacting 1D systems, such as the random-field spin-$1/2$ Heisenberg chain model~\cite{pal_many-body_2010,luitz_many-body_2015}
\be
{\cal H}=\sum_{i=1}^{L}\left(\vec S_i\cdot\vec S_{i+1}-h_iS_i^z\right),
\label{eq:H}
\ee
for which there is now a general 
consensus in the community for an infinite-temperature MBL transition~\cite{luca_ergodicity_2013,luitz_many-body_2015,doggen_many-body_2018,chanda_time_2020,sierant_thouless_2020,abanin_distinguishing_2021}. The very existence of MBL has also been mathematically proven (under minimal assumptions)~\cite{imbrie_many-body_2016} for random interacting Ising chains, and there is a growing number of experimental evidences in 1D~\cite{schreiber_observation_2015,smith_many-body_2016,choi_exploring_2016,roushan_spectroscopic_2017}. 
MBL physics is reasonably well-characterized, mostly thanks to exact diagonalization (ED) techniques~\cite{luitz_many-body_2015,pietracaprina_shift-invert_2018} probing Poisson spectral statistics, low (area-law) entanglement of eigenstates and its out-of-equilibrium logarithmic spreading, eigenstates multifractality. In Fig.~\ref{fig:gs_phase_diag} (c) we show the energy-resolved MBL phase diagram, as obtained in 
Luitz {\it{et al.}}~\cite{luitz_many-body_2015}, for the "standard model" Eq.~\eqref{eq:H}, where $h_i$ are independently drawn form a uniform distribution $[-W_h,W_h]$, and $\epsilon=(E-E_{\rm min})/(E_{\rm max}-E_{\rm min})$ is the energy density above the ground-state.
\subsection{Chapter organization}
The rest of the Chapter will be organized as follows. We start in Sec.~\ref{sec:anderson} with perhaps the simplest case of Anderson localized chains, through the study of the XX spin-1/2 chain model in a random-field. We first briefly discuss its localization properties in real space, and then present numerical (free-fermion) results for the entanglement entropy of many-body (at half-filling) eigenstates, for both the ground-state and at high-energy. Upon varying the intensity of the random-field, we observe interesting scaling behaviors with the localization length, as well as remarkable features in the distribution of von-Neumann entropies. We then move to infinite randomness physics in Sec.~\ref{sec:irfp} with the celebrated logarithmic growth of entanglement entropy for random-bonds XX chains where we unveil an interesting crossover effect, and also for the quantum Ising chain that  is studied at all energies. We then provide a short review of the existing results beyond free-fermions, e.g. random singlet phases with higher spins, and also discuss the cases of engineered disordered systems with locally correlated randomness or the so-called rainbow chain model. We then continue in Sec.~\ref{sec:mbl}
 with the entanglement properties for the many-body localization problem. Eigenstates entanglement entropies at high energy will be discussed for the standard random-field Heisenberg chain model, paying a particular attention to the shape of the distributions in both regimes, and at the transition.
Finally concluding remarks will close this Chapter in Section~\ref{sec:conclusion}.
\section{Entanglement in non-interacting Anderson localized chains}
\label{sec:anderson}
\subsection{Disordered XX chains and single particle localization lengths}
\begin{figure}[t!]
\includegraphics[angle=0,clip=true,width=\columnwidth]{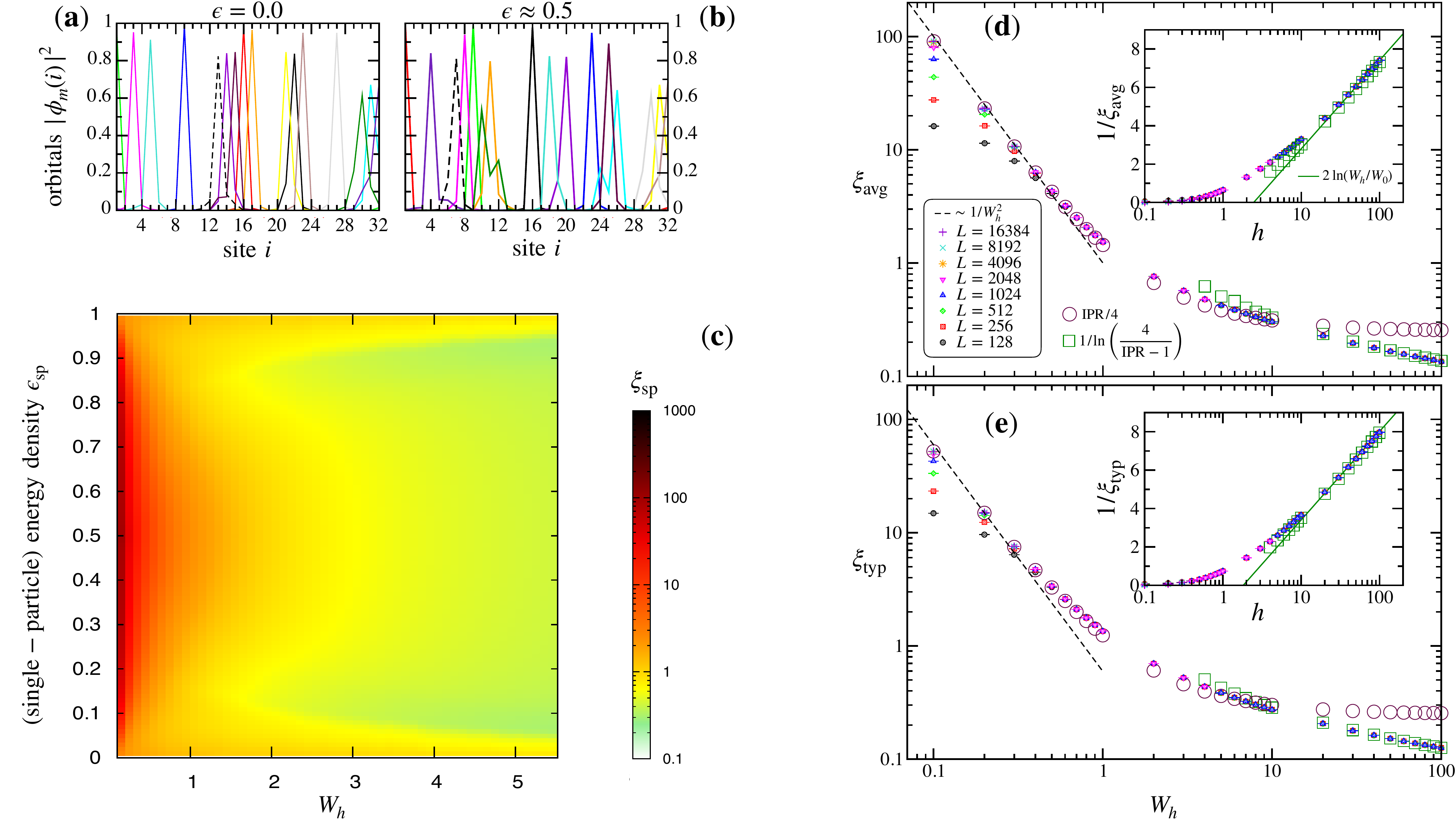}
\caption{Exact diagonalization results for Anderson localization in one dimension Eq.~\eqref{eq:HFF}.  
Panels (a,\,b): exponentially localized orbitals for a single $L=32$ sample (with $W_h=5$). Two exemples of $L/2=16$ occupied orbitals for (a) the many-body ground-state ($\epsilon=0$) and (b) a random high-energy many-body state ($\epsilon\approx 0.5$). Panel (c): color map of the single particle localization length, computed for $L=512$ from the PR Eq.~\eqref{eq:ipr}, and averaged over thousands of disordered samples and small single-particle energy windows. Single-particle energies are normalized, such that for each sample $\epsilon_{\rm sp}=({\cal{E}}-{\cal{E}}_{\rm min})/({\cal{E}}_{\rm max}-{\cal{E}}_{\rm min})$, where ${\cal{E}}$ are the single-particle energies. Panels (d,\,e) show typical and average single-particle localization lengths, computed for various chain lengths $L$ from the PR Eq.~\eqref{eq:ipr} and  averaged over all single particle states and thousands of random independent samples. Limiting cases of large and weak localization lengths are shown with open symbols. Black dased lines show $1/W_h^2$ divergences. Insets: logarithmic divergence of the inverse localization length at strong disorder, Eq.~\eqref{eq:xisd} (green lines) with $W_0\approx 2.46$ for the average (d), and $W_0\approx 1.82$ for the typical (e). Overall, typical and average localization lengths display similar behaviors in the Anderson localized regime.}
\label{fig:anderson}
\end{figure}

Before discussing the entanglement properties,
we first focus on the Anderson localization in real space which occurs in disordered XX chains.
In the easy-plane limit ($\Delta=0$) of Eq.~\eqref{eq:Hs}, the XX chains are equivalent to free fermions
\be
{\cal H}=\sum_{i=1}^{L-1}\left[\frac{J_i}{2}\left(c_i^{\dagger} c_{i+1}^{\vphantom{\dagger}} + c_{i+1}^{\dagger} c_{i}^{\vphantom{\dagger}}\right) -\sum_{i=1}^{L} h_i n_i\right]+{\cal{H_B}},
\label{eq:HFF}
\ee
${\cal{H_B}}$ being a boundary term~\footnote{${\cal{H_B}}=-\frac{J_L}{2}{\rm{e}}^{-i\pi N_f}\left(c_L^{\dagger} c_{1} + c_{1}^{\dagger} c_{L}\right)$ is the the boundary term for PBC (${\cal{H_B}}=0$ for OBC), with $N_f$ the number of fermions ($N_f=S^{z}_{\rm tot}+L/2$).}. This quadratic Hamiltonian takes the diagonal form ${\cal{H}}=\sum_{m=1}^{L}{\cal{E}}_m b_m^{\dagger} b_{m}^{\vphantom{\dagger}}$, using new operators $b_m=\sum_{i=1}^{L}\phi_m(i)c_i$. For non-zero random field, all single particle orbitals $\phi_m(i)$ are exponentially localized in real space, as exemplified in Fig.~\ref{fig:anderson} (a,\,b) for a small chain of $L=32$ sites. 
\subsubsection{Localization length from the participation ratio (PR)} Assuming exponentially localized orbitals $\phi_m$ of the simple (normalized) form
\be
|\phi_m(i)|^2=\tanh\left(\frac{1}{2\xi_m}\right)\exp\left(-\frac{|i-i_0^m|}{\xi_m}\right),
\ee
the participation ratio (PR)~\cite{bell_atomic_1970,edwards_numerical_1972} is given by
\be
{\rm{PR}}_m=\frac{1}{\sum_i|\phi_m(i)|^4}=\frac{\tanh\left(\frac{1}{\xi_m}\right)}{\tanh^2\left(\frac{1}{2\xi_m}\right)}.
\label{eq:ipr}
\ee

In the limit $\xi_m\gg 1$, one recovers the fact the PR is a good estimate of the actual localization length: here ${\rm{PR}}_m\approx 4 \xi_m$. The opposite limit ($\xi\ll 1$) is more tricky. For large disorder $W_h$, a perturbative expansion of the wave function in the vicinity of its localization center $i_0^m$ yields amplitudes vanishing $\sim W_h^{-2r}$, where $r$ is the distance from $i_0^m$. Therefore, for strong randomness,  the localization length slowly vanishes, following
\be 
1/\xi \propto {2\ln W_h} \quad(W_h\gg 1),
\label{eq:xisd}
\ee
and thus $\xi$ can becomes formally smaller (and even much smaller) than the lattice spacing (which has been set to unity). 
However, in the case of a perfectly localized orbital with $\xi\to 0$ the PR will saturate to one, since by definition ${\rm PR}\ge 1$. Therefore, in order to quantify very small localization lengths, one has to slightly modify the way we estimate $\xi_m$. Coming back to the above definition of the PR, Eq.~\eqref{eq:ipr}, $\xi$ will be solution of a cubic equation $X^3+X={2}{{\rm{PR}}}^{-1}$, where $X=\tanh\left(\frac{1}{2\xi}\right)$, thus yielding (using Cardano's formula) 
$X=\left({{\rm PR}^{-1}+\sqrt{\frac{1}{27}+{\rm PR}^{-2}}}\right)^{1/3}+\left({{\rm PR}^{-1}-\sqrt{\frac{1}{27}+{\rm PR}^{-2}}}\right)^{1/3}$.  
At strong disorder (when ${\rm PR}\to 1$) we get $\xi\approx{1}/{\ln\left(\frac{4}{{\rm PR}-1}\right)}$, while in the other limit (${\rm PR_m}\gg 1$) we recover 
$\xi\approx {\rm{PR}}/4$. 

\subsubsection{Numerical results for the localization lengths} Building on Eq.~\eqref{eq:ipr} and the above cubic equation, we have numerically evaluated the average and typical localization lengths for disordered XX chains with constant couplings $J_i=1$ and random fields uniformly distributed in $[-W_h\,,W_{h}]$. In Fig.~\ref{fig:anderson} (d,\,e), we report the disorder dependence  of $\xi_{\rm avg/typ}$, where average is done over all single particle states and $10^4$ independent samples. At weak disorder we observe the expected divergence $\xi\sim 1/W_h^2$~\cite{thouless_relation_1972} while at strong disorder the perturbative result Eq.~\eqref{eq:xisd} is nicely recovered. 
In Fig.~\ref{fig:anderson} (c), the energy-resolved single-particle localization length $\xi_{\rm sp}$ (here averaged over disorder and small energy windows) is shown against $W_h$  as a color map  (collected for $L=512$ sites) where we clearly observe an interesting (albeit weak) delocalization effect at the spectrum edges upon increasing the disorder, a tendency already discussed by Johri and Bhatt~\cite{johri_singular_2012}.

As we will see below, this localization length is an important quantity for the entanglement properties, as $\xi$ will show up in the entanglement entropy.

\subsection{Entanglement entropy for many-body (Anderson localized) eigenstates}
In the non-interacting case, many-body eigenstates are straightforwardly built by filling up a certain number $\nu L$ of single particle states $|m\rangle=b^{\dagger}_m|{\rm{vac.}}\rangle$ (in the following we will work at half-filling $\nu=1/2$). Two types of eigenstates will be considered: the ground-state, occupying the $L/2$ lowest energy states $|{\rm GS}\rangle = \prod_{m=1}^{L/2}b^{\dagger}_m|\rm vac.\rangle$, and  high-energy randomly excited states $|{\rm ES}\rangle = \prod_{m=1}^{L}\theta_m b^{\dagger}_m|\rm vac.\rangle$, where $\theta_m=0$ or $1$, with probability $1/2$ but with the global constraint $\sum\theta_m=L/2$. 

\subsubsection{Free-fermion entanglement entropy}

The  free-fermion entanglement entropy of a subsystem ${\cal {A}}$ ($i=1,\ldots, \ell \in {\cal{A}}$)  is easy to compute~\cite{peschel_reduced_2004} using the $\ell\times \ell$ correlation matrix
${\mathcal{C}}_{\cal A}$, defined by
\begin{equation}
\label{eq:C}
{\mathcal{C}}_{\cal A}~=~\begin{pmatrix}\langle c^{\dagger}_{1}c_1^{\vphantom{\dagger}}\rangle & \langle c^{\dagger}_{1}c_2^{\vphantom{\dagger}}\rangle  & \cdots &  \langle c^{\dagger}_{1}c_{\ell}^{\vphantom{\dagger}}\rangle \\
\langle c^{\dagger}_{2}c_1^{\vphantom{\dagger}}\rangle & \langle c^{\dagger}_{2}c_2^{\vphantom{\dagger}}\rangle & \ddots & \vdots\\
\vdots &  & \ddots &  \\
 & &  & \langle c^{\dagger}_{\ell}c^{\vphantom{\dagger}}_{\ell}\rangle
\end{pmatrix}
,
\end{equation}
with matrix elements $\langle c^{\dagger}_{i}c_j^{\vphantom{\dagger}}\rangle$ evaluated in a given many-body (ground or excited) eigenstate. 
The von-Neumann entanglement entropy is then given by
\be
{{S}}_{\rm vN}=-\sum_{n}\Bigl[\lambda_n \ln \lambda_n +(1-\lambda_n) \ln
(1-\lambda_n)\Bigr],
\label{eq:S}
\ee
where the $\lambda_n$ are the eigenvalues of ${\mathcal{C}}_{\cal A}$.
\begin{figure}[t!]
\includegraphics[angle=0,clip=true,width=\columnwidth]{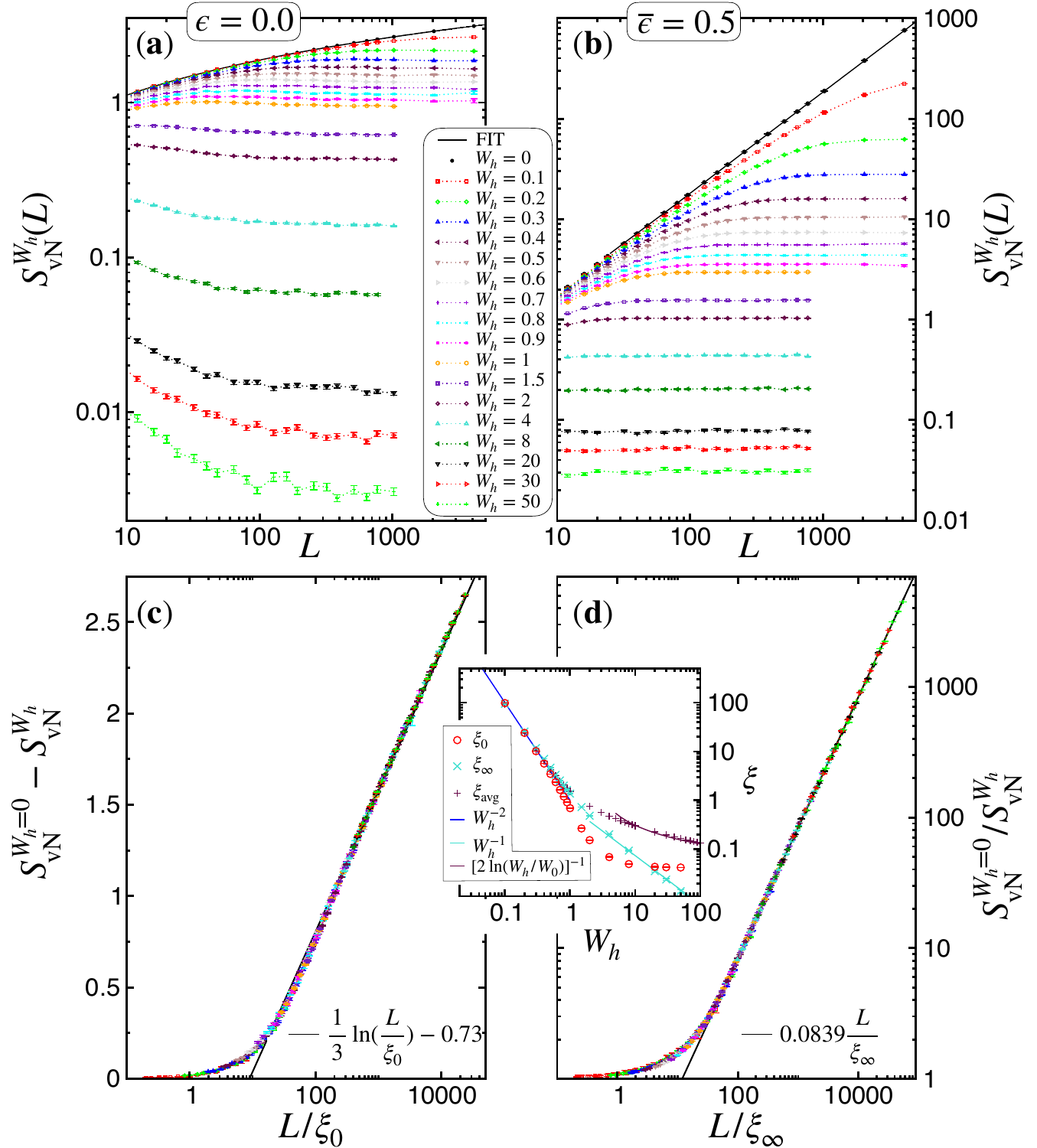}
\caption{Von-Neumann entanglement entropies (with half-chain entanglement cuts) for  Anderson localized chains [periodic XX chains in a random field, Eq.~\eqref{eq:HFF}]: ED results for ground-state ($\epsilon=0$, left panels) and randomly excited states (${\overline{\epsilon}}=0.5$, right panels), averaged over thousands of independent disordered samples. (a,\,b) Entanglement scaling $S_{\rm vN}^{W_h}(L)$ shown for various disorder strengths $W_h$. The black lines are fits to the clean case forms: Eq.~\eqref{eq:cc} with $c=1$ and $\rm constant =0.344$ for the ground-state (a) ; and Eq.~\eqref{eq:clean} with $s_0\approx 0.1845$, and $\rm constant\approx -0.476$ for high energy (b). Panels (c) and (d) show data collapses using the scaling forms Eq.~\eqref{eq:scaling} and Eq.~\eqref{eq:vl}. Lines are the asymptotic forms, indicated on the plots. Inset: Log-log plot of the disorder dependence of the different length scales : $\xi_0$ and $\xi_\infty$ are shown together with the average single-particle localization length $\xi_{\rm avg}$ (see also Fig.~\ref{fig:anderson}). They all display the same $W_h^{-2}$ divergence at weak disorder, while the strong disorder behavior is non-universal (see text).}
\label{fig:S_anderson}
\end{figure}

\subsubsection{Low and high energy}
\paragraph*{(i) Zero temperature---} In the absence of disorder, the $T=0$ entanglement entropy of a periodic XX chain follows the famous log scaling~\cite{vidal_entanglement_2003,calabrese_entanglement_2004,korepin_universality_2004}
\be
S_{\rm vN}^{W_h=0}(\ell)=\frac{c}{3}\ln \ell +\rm constant,
\label{eq:cc}
\ee
with the central charge $c=1$. For Anderson localized chains, the log growth is cutoff by the finite localization length, as clearly visible in Fig.~\ref{fig:S_anderson} (a) for periodic systems with a half-chain entanglement cut ($\ell=L/2$). Perhaps more interestingly, the following scaling behavior emerges
\bea
S_{\rm vN}^{W_h=0}-S_{\rm vN}^{W_h}&\propto&\frac{1}{3}\ln\left(\frac{L}{\xi_0}\right)\quad{\rm{if}}\quad L\gg \xi_0\label{eq:scaling}\\
&\to &0\quad\quad\quad\quad\quad\,{\rm{if}}\quad L\ll \xi_0,
\eea
as visible in panel (c) of Fig.~\ref{fig:S_anderson}. The extracted length scale $\xi_0$ is plotted against $W_h$ in the inset of Fig.~\ref{fig:S_anderson} together with the average single-particle localization length $\xi_{\rm avg}$ (also previously shown in Fig.~\ref{fig:anderson}). The $W_h^{-2}$ divergence at weak disorder is clearly observed, while at stronger disorder the behavior is non-universal (see below for a discussion).\\

\paragraph*{(ii) Infinite temperature---}The high-energy case is also very interesting, see Fig.~\ref{fig:S_anderson} (b,\,d). In the absence of disorder the following volume-law entanglement entropy is observed 
\be
S_{\rm vN}^{W_h=0}(\ell)=s_0 L+\rm constant,
\label{eq:clean}
\ee
with a volume-law coefficient $s_0\approx 0.1845$, which clearly departs from Page's law~\cite{page_average_1993} (as clearly understood in Ref.~\cite{vidmar_entanglement_2017}), and an additive $\rm constant\approx -0.476$. As for the zero-temperature situation, as soon as $W_h\neq 0$ Anderson localization leads to the saturation of the von-Neumann entropy, even  at infinite temperature. In addition, we also observe in Fig.~\ref{fig:S_anderson} (d) a scaling behavior for
\bea
S_{\rm vN}^{W_h=0}/S_{\rm vN}^{W_h}&\sim&{L}/{\xi_\infty}\quad{\rm{if}}\quad L\gg \xi_{\infty}\label{eq:vl}\\
&\to &1\quad\quad\quad{\rm{if}}\quad L\ll \xi_\infty.
\eea
The extracted length scale $\xi_\infty$, visible in Fig.~\ref{fig:S_anderson} (inset), also diverges $\sim W_h^{-2}$ at weak disorder, and equally coincides with $\xi_0$ and $\xi_{\rm avg}$.

\subsubsection{Strong disorder limit}
\label{sec:SD}
It is worth briefly discussing the strong disorder situation, which may also be relevant for the MBL problem (see Section~\ref{sec:mbl}). Despite their similar weak disorder properties, the three length scales $\xi_{\rm avg}$, $\xi_0$ and $\xi_{\infty}$ (inset of Fig.~\ref{fig:S_anderson}) display distinct behaviors at strong $W_h$, and neither $\xi_0$ nor $\xi_\infty$ shows the logarithmic divergence Eq.~\eqref{eq:xisd} of $\xi_{\rm avg}$. This is in fact easy to understand from the strong disorder limit of $S_{\rm vN}$.\\

\begin{figure}[t!]
\includegraphics[angle=0,clip=true,width=\columnwidth]{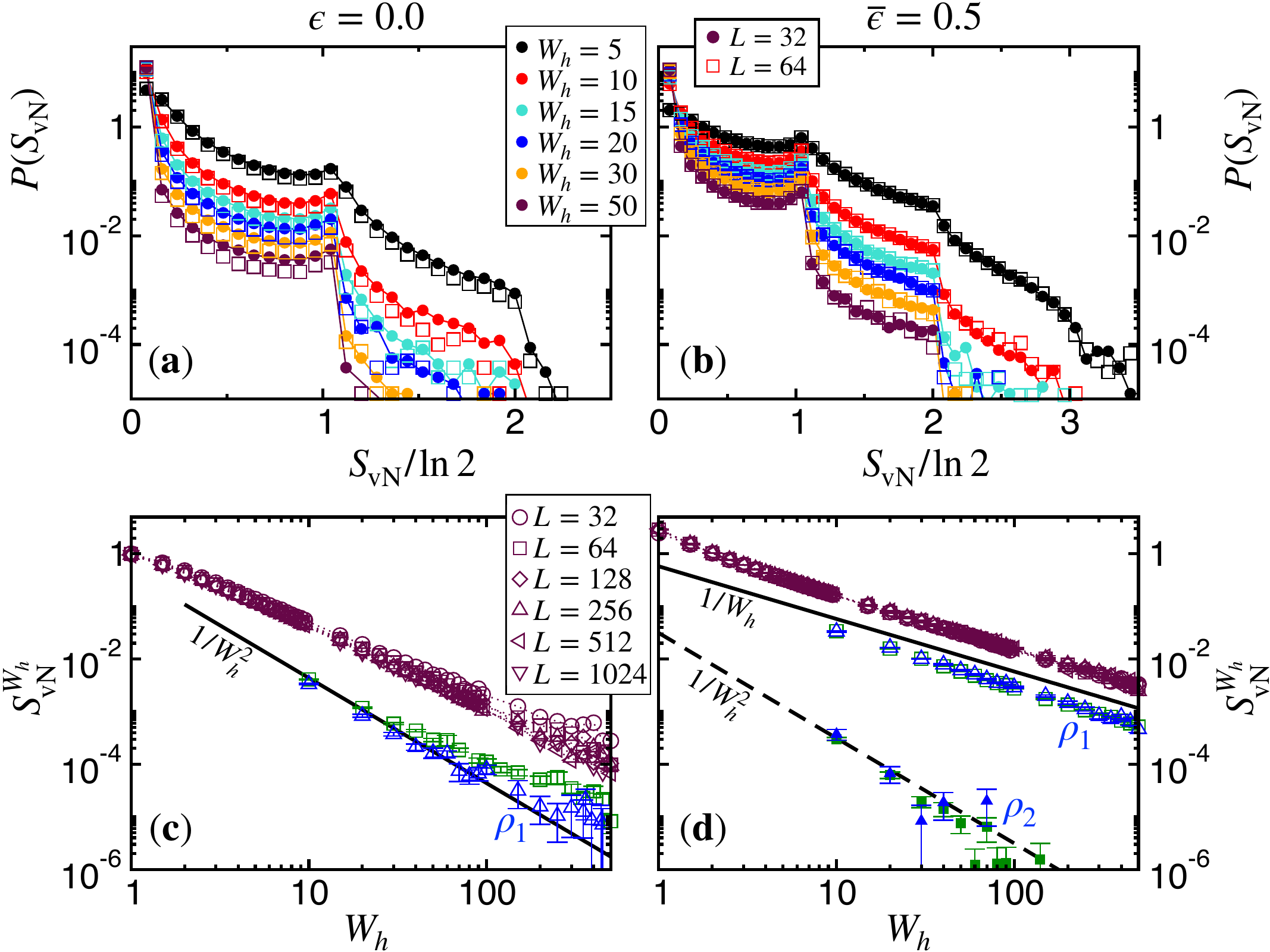}
\caption{Strong disorder behavior of the half-chain entanglement entropies for Anderson localized chains. ED results for ground-state ($\epsilon=0$, left panels) and randomly excited states (${\overline{\epsilon}}=0.5$, right panels).
Top panels (a,\,b) show histograms of $S_{\rm vN}/\ln 2$  collected for $L=32,\,64$ over several hundred thousands of independent random samples for varying disorder strengths, as indicated on the plot. One clearly sees significant peaks at integer values, thus signalling anomalously weak disordered sites at the entanglement cut (see text).
Bottom panels (c,\,d) show the strong disorder behavior of the average entropy, consistent with power-law decay (see text) with distinct exponents between ground and excited states. Note also the strong finite size effects at $\epsilon=0$ are almost absent at high energy. The strong disorder scaling of $S_{\rm vN}$ is dominated by the probability $\rho_1=P(|S_{\rm vN}/\ln 2-1|\le 0.05)$.}
\label{fig:ps_anderson}
\end{figure}

\paragraph*{(i) Ground-state---}At $T=0$ the average is dominated by rare singlet pairs yielding $S_{\rm vN}=\ln 2$,  appearing only if two neighbors have weak disorder, which occurs with a very low probability $\sim 1/W_h^2$. We therefore expect for the strong disorder average entropy $S_{\rm vN}\sim W_h^{-2}$, and hence a non-vanishing localization length $\xi\sim \exp\left(AW_h^{-2}\right)$, even at very strong disorder. This simple argument can be numerically confirmed. In Fig.~\ref{fig:ps_anderson} (a) the histograms $P(S_{\rm vN})$ clearly show a peaked structure with a dominant peak at $0$ and a secondary one at $\ln 2$. This is further checked in Fig.~\ref{fig:ps_anderson} (c) where the disorder-average entanglement entropy, together with the probability to observe $\ln 2$, $\rho_1=P(|S_{\rm vN}/\ln 2-1|\le 0.05)$ both show a clear $W_h^{-2}$ decay at large disorder, thus validating the above scenario. Note that non-negligible finite size effects are present for the ground-state, while randomly excited states (discussed below) shown in panels (b,\,d) are much less spoiled by finite chain effects.\\

\paragraph*{(ii) Excited-states---}In the same spirit, one can also make some predictions for the high-energy behavior. Indeed, at high temperature thermalization is expected for each individual site having locally a weak disorder, which occurs with a higher probability $\sim 1/W_h$. 
We therefore expect $S_{\rm vN}\sim W_h^{-1}$ at strong disorder, thus implying that $\xi_\infty\sim W_h^{-1}$, a behavior nicely observed in Fig.~\ref{fig:S_anderson} (inset). Again such a simple strong disorder argument is numerically confirmed in Fig.~\ref{fig:ps_anderson} (b) where the histograms $P(S_{\rm vN})$ also have a peaked structure with a dominant peak at $0$ and a richer secondary peak arrangement, with one at $\ln 2$ and another visible at $2\ln 2$. This is further checked in panel (d) where the disorder-average von-Neumann entropy, together with the probability $\rho_1$, both display a nice $W_h^{-1}$ decay at large disorder, almost size-independent contrasting with the ground-state. The third peak at $2\ln 2$ can also be tracked with $\rho_2=P(|S_{\rm vN}/2\ln 2-1|\le 0.05)$, which agrees with a $\sim W_h^{-2}$ decay, while it reaches the limit of numerics.

\newpage\section{Entanglement and infinite randomness criticalities}
\label{sec:irfp}
In the context of random quantum magnets, the strong disorder renormalization group (SDRG)  method~\cite{ma_random_1979,fisher_critical_1995,igloi_strong_2005} have proven to be very useful, in particular for the celebrated infinite randomness fixed point (IRFP) physcis, which has been deeply described by D. S. Fisher in a series of seminal papers for $d=1$~\cite{fisher_random_1992,fisher_random_1994,fisher_critical_1995}, then later extended to $d>1$~\cite{motrunich_infinite-randomness_2000,lin_low-energy_2003,kovacs_infinite-disorder_2011}, and applied to a broad range of systems~\cite{igloi_strong_2005,igloi_strong_2018}.
\subsection{Entanglement in disordered XXZ and quantum Ising chains}
\subsubsection{Random singlet state for disordered $S=1/2$ chains} 
Building on the SDRG framework for random-exchange antiferromagnetic XXZ chains~\cite{fisher_random_1994} (Eq.~\eqref{eq:Hs} with $h_i=0$), or for the random $d=1$ TFIM~\cite{fisher_critical_1995} at criticality (Eq.~\eqref{eq:TFI} with $\delta={\overline{\ln J}}-{\overline{\ln h}}=0$), Refael and Moore~\cite{refael_entanglement_2004,refael_criticality_2009} have shown that infinite randomness criticality is accompanied by a logarithmic scaling for the disorder-average entanglement entropy, of the form
\be
S(\ell)=\frac{c_{\rm eff}}{3}\ln \ell +\rm constant,
\label{eq:rsp}
\ee
thus contrasting with the previously discussed Anderson localization case where $S$ is bounded by the finite localization length. In the above form, the coefficient $c_{\rm eff}=c\ln 2$ has been reduced by a factor $\ln 2$  as compared to the disorder-free (conformally invariant) case in Eq.~\eqref{eq:cc}. This result is a  direct consequence of the random-singlet structure of the ground-state of the random XXZ chain~\cite{fisher_random_1994} where the probability to form a singlet between two sites at distance $\ell$ is $\propto \ell^{-2}$~\cite{fisher_random_1994,hoyos_correlation_2007} (see also the recent work by Juh\'asz~\cite{juhasz_corrections_2021} for a SDRG analysis of subleading corrections).\\

\begin{figure}[t!]
\begin{center}
\includegraphics[width=\columnwidth,clip]{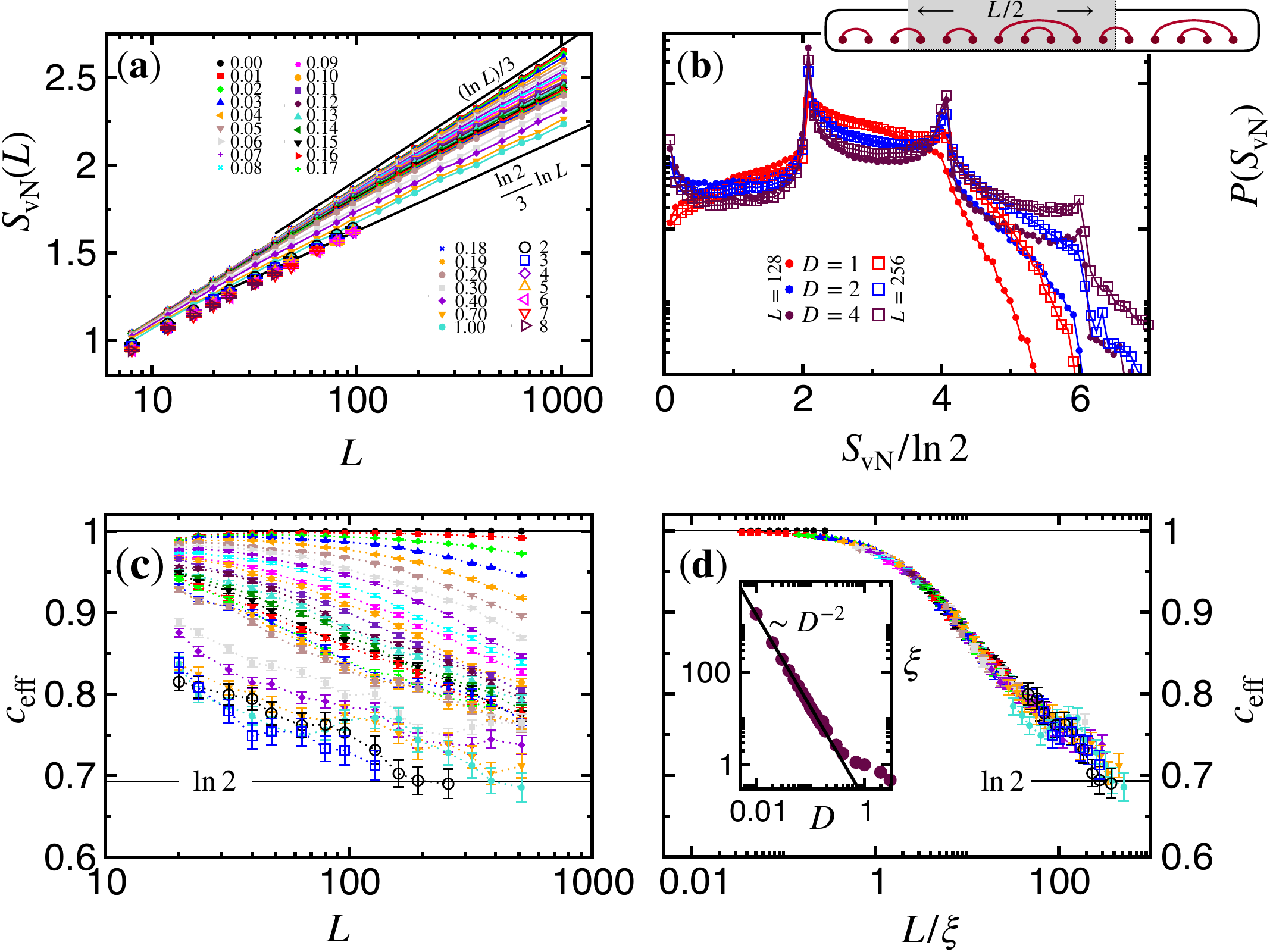}
\caption{Exact diagonalization results for the (half-chain) ground-state  von-Neuman entropy of $S=1/2$ XX chains with random bonds Eq.~\eqref{eq:randomxx} with power-law distributed couplings $P(J)\propto J^{-1+1/D}$, averaged over several thousands of disordered samples. (a) Logarithmic scaling Eq.~\eqref{eq:rsp} shown for many disorder strength $D=0,\ldots,\,8$ (indicated on the plot) thus emphasizing the crossover between clean and RSP behaviors. The prefactor $c_{\rm eff}$ of the logarithmic growth, extracted from fits to the form Eq.~\eqref{eq:rsp} over sliding 7-point windows, is shown for $D\in [0.01,\,3]$ in panel (c), and by rescaling the system size $L\to L/\xi$ in panel (d) where a reasonable data collapse is obtained. Inset (d): the extracted crossover length scale diverges $\sim D^{-2}$ (we have fixed $\xi=1$ for $D=1$). Panel (b) shows histograms of $S_{\rm vN}/\ln 2$  collected for $L=128,\,256$ over several hundred thousands of independent random samples for varying disorder strengths, as indicated on the plot. The random singlet structure (see also the schematic picture on top right) clearly develops upon increasing disorder and/or system size, with peaks at even integer values.}
\label{fig:RSP}
\end{center}
\end{figure}

\paragraph*{(i) Large-scale numerics for random XX chains---}
The SDRG analytical prediction Eq.~\eqref{eq:rsp} with $c_{\rm eff}=c\ln 2$ has been numerically confirmed using free-fermion exact diagonalization calculations at the XX point for large chains~\cite{laflorencie_scaling_2005,hoyos_correlation_2007,igloi_finite-size_2008,fagotti_entanglement_2011,pouranvari_entanglement_2013}. Here in this work, we will discuss new numerical results (see Fig.~\ref{fig:RSP}) for random XX chains, governed by
\be
{\cal{H}}_{\rm random~XX}=\sum_{i=1}^{L}J_i\left(S_{i}^xS_{j}^x+S_{i}^yS_{j}^y\right)
\label{eq:randomxx}
\ee
with power-law distributed AF couplings $P(J)\propto J^{-1+1/D}$. Note that such a distribution allows to describe a broad range of disorder strengths: from clean physics $D\to0$ to the  infinite randomness fixed point distribution where $D\to \infty$.

Fig.~\ref{fig:RSP} (a) shows the finite size behavior of the disorder-average von-Neumann entropy $S_{\rm vN}(L)$ (here again we focus on half-chain cuts), for a broad range of initial disorder strengths $D=0.01,\ldots,\,8$. We clearly observe the logarithmic scaling Eq.~\eqref{eq:rsp}, with a smooth finite-size crossover from clean physics $c_{\rm eff}=1$ to the SDRG asymptotic result~\cite{refael_entanglement_2004} $c_{\rm eff}=\ln 2$ observed at large enough $D$ or $L$.\\

\paragraph*{(ii) Crossover phenomenon---} This crossover is controlled by a disorder-dependent  length scale $\xi$, as  studied in panels (c,\,d). There,  the prefactor of the logarithmic growth has been extracted from simple fit to the form Eq.~\eqref{eq:rsp} over sliding windows containing 7 points. The disorder and size dependent crossover for the "effective central charge" $c_{\rm eff}(L,D)$ (between 1 and $\ln 2$), exhibits a "universal" scaling form $c_{\rm eff}(L/\xi)$, as extracted in Fig.~\ref{fig:RSP} (d). Moreover, $\xi$ plotted in panel (d) inset is found to diverge $\propto D^{-2}$ at weak disorder. This remarkable behavior is in perfect agreement with a crossover already identified for the average correlation functions~\cite{laflorencie_comment_2003,laflorencie_crossover_2004,laflorencie_scaling_2004}. As a matter of fact, $\xi$ gives a simple quantitative scale beyond which asymptotic results from SDRG can be expected. For instance, on finite chains the random singlet structure (depicted in Fig.~\ref{fig:RSP} (b) inset) becomes effectively visible, either when the initial disorder $D$ is strong enough, or for increasing system size, as clearly visible in Fig.~\ref{fig:RSP} (b).\\

\paragraph*{(iii) Random singlets: significant others ---} The situation is also verybinteresting for higher R\'enyi indices, as discussed by Fagotti {\it{et al.}}~\cite{fagotti_entanglement_2011}. 
Depending on how the averaging over disorder is performed, one should expect the different scalings
\bea
{S}_q&=&\frac{{\overline{\ln {\rm Tr}\rho_A^q}}}{1-q}=\frac{\ln 2}{3}\ln L+{\rm{const}}_q,\\
{\tilde{~S_q~}}&=&\frac{\ln{\overline{ {\rm Tr}\rho_A^q}}}{1-q}=f_q\frac{\ln 2}{3}\ln L+{\rm{const}}_{q}^{'},
\label{eq:Sqtilde}
\eea
with the non-trivial prefactor $f_q=\frac{3\left(\sqrt{5+2^{3-q}}-3\right)}{2\ln 2 (1-q)}\le 1$, vanishing at large $q$ and $f_q\to 1 $ in the von-Neumann (or Shannon) limit $q\to 1$. This peculiar dependence on the disorder averaging is one of the hallmark of infinite randomness physics, as deeply discussed by D. S. Fisher for correlations functions~\cite{fisher_random_1994,fisher_critical_1995}.

It is also worth mentioning how the works on entanglement in the RSP (given by a rather simple counting of singlet bonds crossing the entanglement cut) led to the emergence of the idea of a valence bond entanglement entropy~\cite{alet_valence_2007,chhajlany_topological_2007,
mambrini_hard-core_2008,jacobsen_exact_2008,alet_valence-bond_2010,tran_valence_2011}. This alternative entanglement witness turns out to be much easier to access within quantum Monte Carlo frameworks, as compared to the von-Neumann or R\'enyi entanglement entropies~\cite{hastings_measuring_2010,kallin_anomalies_2011,humeniuk_quantum_2012,
helmes_entanglement_2014,luitz_improving_2014,kulchytskyy_detecting_2015,toldin_entanglement_2019}, despite some recent impressive progresses~\cite{demidio_entanglement_2020,francesconi_strong_2020}.

Random singlet physics has also recently triggered new studies, such as the investigation of the entanglement negativity in Refs.~\cite{ruggiero_entanglement_2016,turkeshi_negativity_2020}, or the extension of the concept of symmetry-resolved entanglement equipartition~\cite{laflorencie_spin-resolved_2014,goldstein_symmetry-resolved_2018,xavier_equipartition_2018,murciano_symmetry_2020} to the RSP by Turkeshi {\it{et al.}} in Ref.~\cite{turkeshi_entanglement_2020}.

\subsubsection{Infinite randomness criticality at high energy}
\begin{figure}[b!]
\begin{center}
\includegraphics[width=.85\columnwidth,clip]{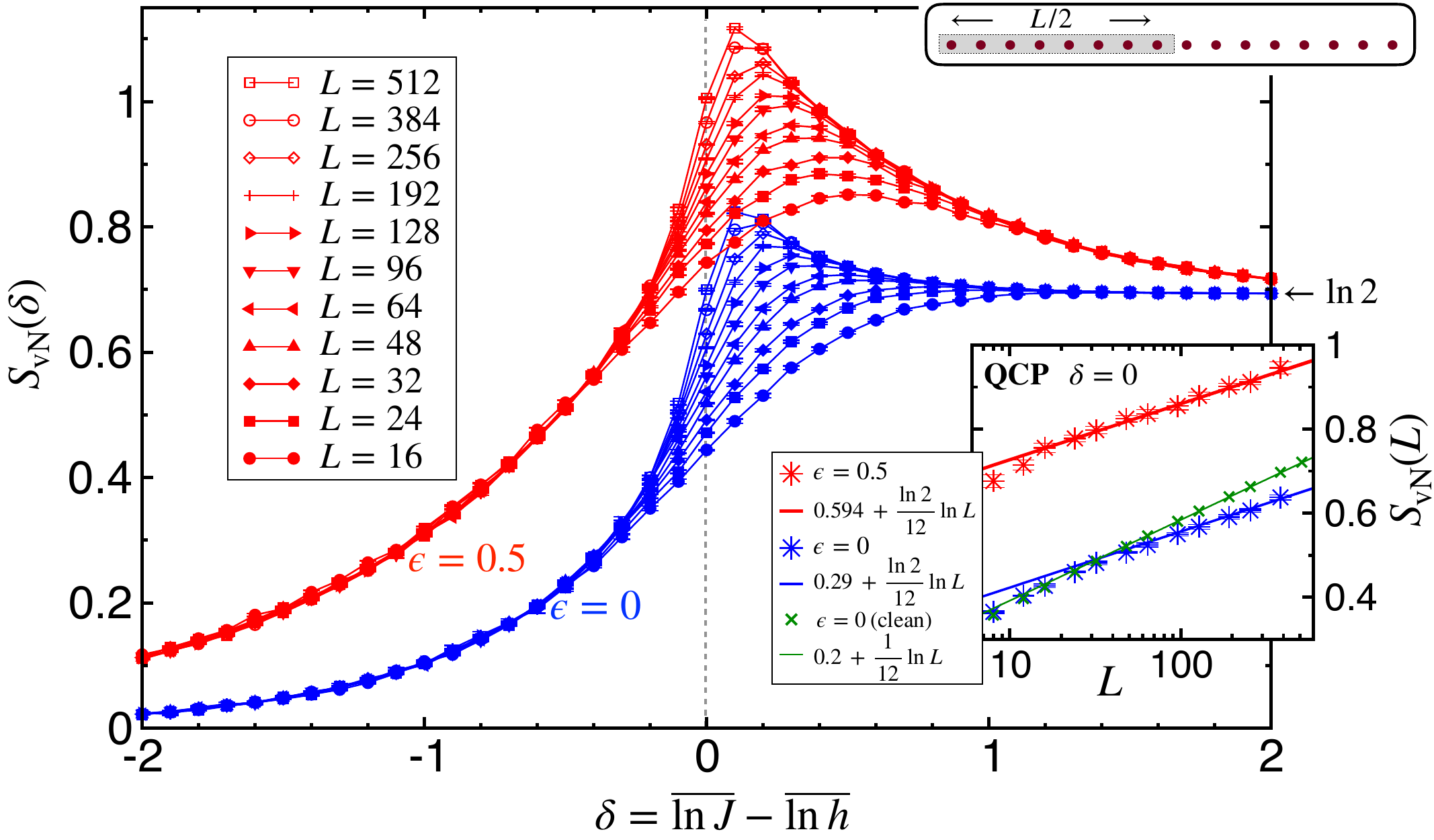}
\caption{Exact diagonalization results for the random TFIM Eq.~\eqref{eq:TFI} with open boundary conditions. Results are averaged over several thousands of samples for various system lengths $L$, as indicated on the plot. The half-chain von-Neumann entropy (see schematic picture, top right), is plotted against the control parameter $\delta$ for (zero-temperature) ground-state ($\epsilon=0$, blue symbols) and infinite-temperature ($\epsilon=0.5$,  red symbols), in both cases showing qualitatively similar behaviors (see text). Inset: the critical scaling at $\delta=0$ takes the expected logarithmic form Eq.~\eqref{eq:logTFI}. Note also the crossover between the clean case ($\epsilon=0$, green symbols) and the asymptotic behavior.}
\label{fig:TFIM}
\end{center}
\end{figure}

As expected from high-energy SDRG approaches~\cite{pekker_hilbert-glass_2014,vasseur_quantum_2015,you_entanglement_2016,monthus_strong_2018}, the zero-temperature quantum criticality of the disordered quantum Ising chain Eq.~\eqref{eq:TFI} must remain unchanged at {\it{all}} energies, so far only confirmed by a single numerical study~\cite{huang_excited-state_2014}. Here we present and discuss our numerical results obtained for the 1D random TFIM in  Fig.~\ref{fig:TFIM}. 
First, at criticality when $\delta={\overline{\ln J}}-{\overline{\ln \,h}}=0$, we check in the inset of Fig.~\ref{fig:TFIM} the logarithmic scaling for the disorder-average entropy with open boundary conditions with a cut at half-chain (see schematic picture in Fig.~\ref{fig:TFIM}, top right) 
\be
S_{\rm vN}(L/2,\epsilon,\delta=0)=\frac{\ln 2}{12} \ln L +\rm const(\epsilon),
\label{eq:logTFI}
\ee
where the only dependence on the energy density $\epsilon$ comes in the non-universal additive constant. We remind that ground-state is at $\epsilon=0$, while $\epsilon=0.5$ corresponds to infinite-temperature states. Interestingly, we also remark that ${\rm const}(0.5)\approx 2\times {\rm const}(0)$. In a way similar to the previously discussed crossover from clean to IRFP for the random-bod XX chain, we also observe the same effect here.
However we will not vary the disorder strength, but instead vary the control parameter $\delta={\overline{\ln J}}-{\overline{\ln \,h}}=2\ln W$, keeping couplings and fields drawn from box distributions: $P_{J/h}={\rm{Box}}[0\,,W_{J/h}]$ uniform between $0$ and $W_{J/h}$, with $W_J=W_{h}^{-1}=W$.

In the main panel of Fig.~\ref{fig:TFIM},  upon varying $\delta$ the von-Neumann entropy displays qualitatively similar behaviors for zero and infinite temperature: (i) area-law entanglement, even at high temperature ; (ii) $S_{\rm vN}\to \ln 2$ for positive $\delta$, signaling localization protected quantum-order~\cite{huse_localization-protected_2013} with a "cat-state" structure for the eigenstates ; (iii) IRFP log scaling Eq.~\eqref{eq:logTFI} at criticality (see inset).

\subsection{Other systems showing infinite randomness criticality}
\subsubsection{Higher spins, golden chain, and RG flows}
Back to zero-temperature, infinite randomness physics also occurs for higher spin systems with $S>1/2$ chains~\cite{hyman_impurity_1997,monthus_phases_1998,refael_spin_2002,damle_permutation-symmetric_2002}, for which it was shown~\cite{refael_entanglement_2007,saguia_entanglement_2007} that 
\be
S_{\rm vN} = \frac{\ln(2S+1)}{3}\ln L +{\rm constant}.
\ee
Non-abelian RSP are also expected for disordered chains of Majorana or Fibonacci anyons~\cite{bonesteel_infinite-randomness_2007,fidkowski_textitc_2008,fidkowski_permutation-symmetric_2009}, with a logarithmic von-Neumann entropy whose "effective central charge" pre-factor is given by $\ln {\cal{D}}$, where ${\cal{D}}$ is the quantum dimension, {\it{e.g.}} ${\cal{D}}={\sqrt{2}}$ for a Majorana chain (quantum Ising chain at criticality), and ${\cal{D}}=(1+\sqrt{5})/2$ for Fibonacci anyons.

\begin{figure}[h!]
\begin{center}
\includegraphics[width=0.45\columnwidth,clip]{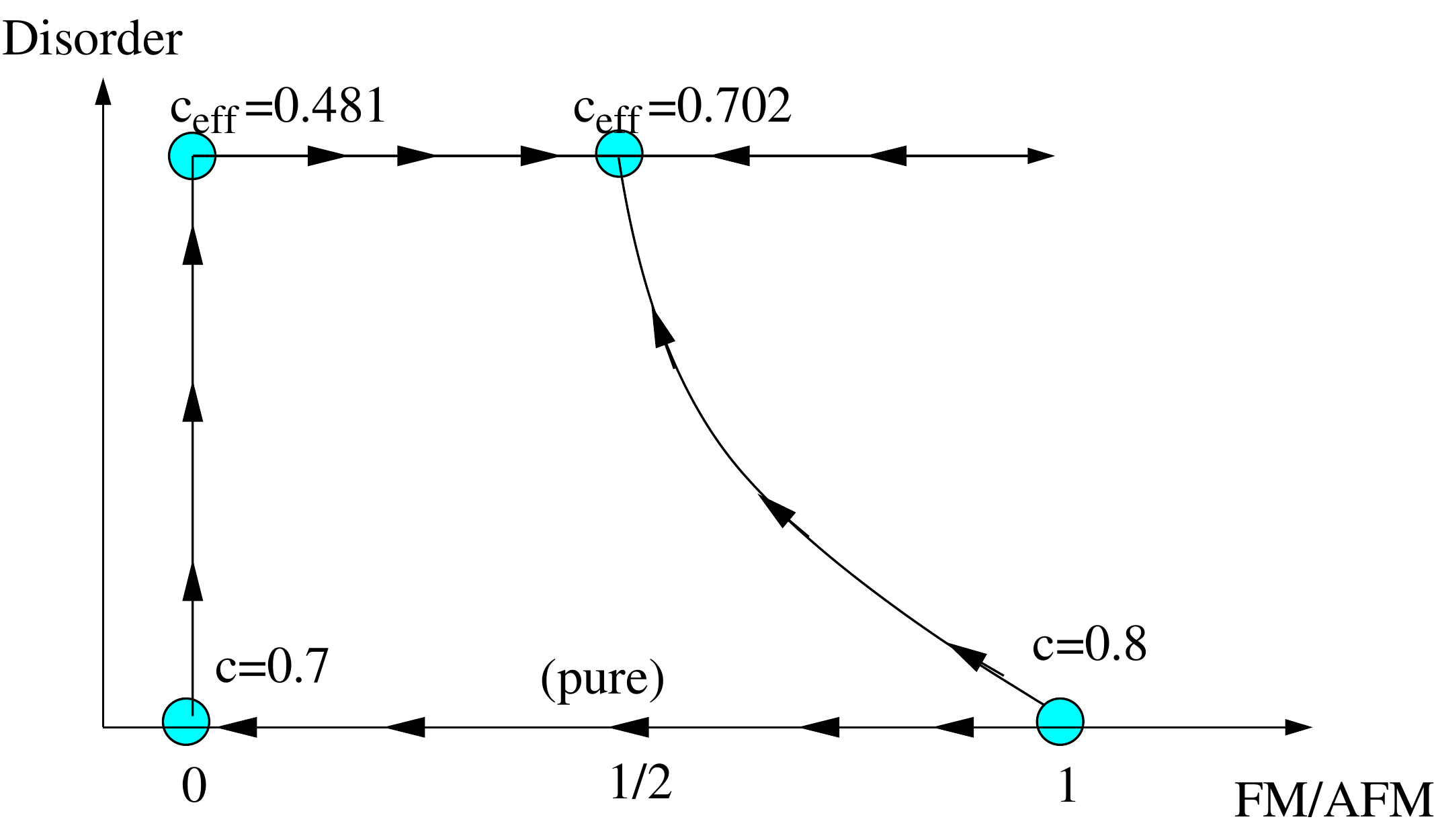}
\caption{RG flow diagram of pure and random golden chains. In the clean case the Zamolodchikov's $c$-theorem is verified, while this is not necessarily true in the disordered case. Figure taken from Fidkowski {\it{et al.}}~\cite{fidkowski_textitc_2008}.}
\label{fig:golden}
\end{center}
\end{figure}

There is an important issue concerning the entanglement gradient along RG flows. In the absence of disorder, the famous Zomolodchikov's $c$-theorem~\cite{zomolodchikov_``irreversibility_1986} implies a decay of entanglement along RG flows. The observation of decreasing entropies along infinite randomness RG flows~\cite{refael_entanglement_2004,laflorencie_scaling_2005,refael_entanglement_2007} then raised a similar question for random systems. However two clear counter examples have ruled out such a scenario, due to Santachiara~\cite{santachiara_increasing_2006} for generalized quantum Ising chains including the $N$-states random Potts chain, and later by Fidkowski {\it{et al.}}~\cite{fidkowski_textitc_2008} for disordered chains of Fibonacci anyons. The RG flow phase diagram of disordered golden chains from Ref.~\cite{fidkowski_textitc_2008} is given in Fig.~\ref{fig:golden}, see also Ref.~\cite{refael_criticality_2009}.

\subsubsection{$d>1$ Infinite randomness}
Infinite randomness physics is not restricted to $d=1$, but also occurs for $d\ge 2$ random quantum Ising models~\cite{motrunich_infinite-randomness_2000,kovacs_renormalization_2010,kovacs_infinite-disorder_2011,monthus_random_2012}, while $d>1$ random-exchange antiferromagnets do not host random singlet physics since the $T=0$ N\'eel order is very robust against disorder~\cite{lin_low-energy_2003,laflorencie_random-exchange_2006}.

There has been some controversy regarding the precise scaling of the von-Neumann  entropy for higher dimensional IRFP in the random TFIM, in particular for the $d=2$ square lattice~\cite{lin_entanglement_2007,yu_entanglement_2008}. Building on an improved SDRG algorithm~\footnote{In Refs.~\cite{kovacs_renormalization_2010,kovacs_infinite-disorder_2011} the ${\cal O}(N^3)$ CPU time scaling of the simplest SDRG approaches was scaled down to ${\cal O}(N \ln N)$ for arbitrary dimension, allowing to study the entanglement of $d=2,3,4$ disordered quantum Ising models up to $N\sim 10^6$ spins, see also Ref.~\cite{kovacs_universal_2012}.}, Kov{\'a}cs and Igl{\'o}i~\cite{kovacs_renormalization_2010,kovacs_infinite-disorder_2011} unambiguously found a pure area-law scaling with additive (negative) logarithmic corrections~\cite{yu_entanglement_2008,kovacs_universal_2012}, coming from the subsystem corners:
\be
S_{\rm vN}=\alpha L+4\ell_{1}(\pi/2)\ln L +{\rm const.}
\ee
with $\ell_{1}(\pi/2)\approx -0.03$. These logarithmic corrections, induced by sharp subsystem boundaries, only occur at the infinite randomness criticality~\cite{kovacs_universal_2012}. Interestingly, they are of the same order of magnitude as the corner terms which show up in (disorder-free) $2+1$ CFT~\cite{bueno_universality_2015,bueno_universal_2015,bueno_corner_2015}.

\subsection{Engineered disorders}
In this part we discuss a class of disordered spin chains where some local correlations have been included, thus making the systems not entirely random. Two main examples will be addressed: (i) a simple TFIM with purely local correlations between random couplings and fields~\cite{binosi_increasing_2007,hoyos_protecting_2011}, and (ii) the so-called "rainbow model" introduced in Ref.\cite{vitagliano_volume-law_2010}, and its subsequent extensions.\\

\begin{figure}[hb!]
\begin{center}
\includegraphics[width=.85\columnwidth,clip]{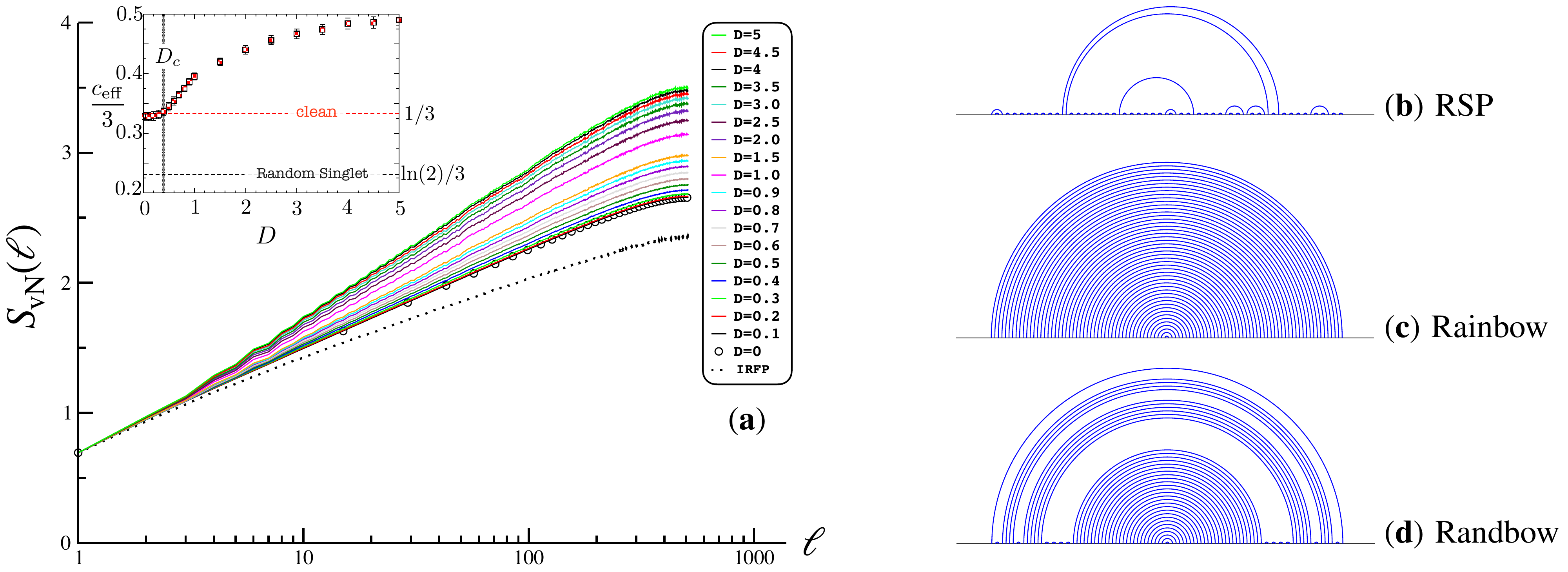}
\caption{Left (a): Random TFIM with local correlations Eq.~\eqref{eq:Hcorrelated}. Disorder-averaged entanglement (von-Neumann) entropy plotted against subsystem lengths $\ell$ for the ground-state of Eq.~\eqref{eq:Hcorrelated} with various disorder strengths $D$, couplings being power-law distributed $P(J)\propto J^{-1+1/D}$, for $L=1024$ sites and $5000$ disorder realizations. Inset: the disorder dependence of the effective central charge exhibits a transition for $D_c\approx 0.3$. Figure taken from  Ref.~\cite{hoyos_protecting_2011}. Right: Sketch of the three regimes of the randbow chain model Eq.~\eqref{eq:randbow} with randomness in the couplings $J_i$. (b) RSP when $h\ll 1$ ; (c) Rainbow phase $h\to \infty$ ; (d) Randbow regime $h\sim 1$. Figure taken from  Ref.~\cite{alba_unusual_2019}.}
\label{fig:correlated}
\end{center}
\end{figure}

\paragraph*{(i) Random quantum Ising chains with locally correlated disorder---}
Binosi {\it{et al.}}~\cite{binosi_increasing_2007} first proposed the following quantum Ising chain model with a very simple purely local correlation in the disorder parameters:
\be
{\cal{H}}=-\sum_i J_i\left(\sigma_i^x \sigma_{i+1}^x +\sigma_i^z\right),
\label{eq:Hcorrelated}
\ee
as an exemple which exhibits growing entanglement upon increasing disorder.  In the above Hamiltonian, it is remarkable to see that the very same (random) number $J_i$ acts on a site $i$ as a field as well as a coupling on its adjacent bond, such that a perfect correlation (while purely local, with a minimal correlation length) is achieved. Building on field theory, SDRG, and free-fermion numerics, this model was studied by Hoyos {\it{et al.}} in Ref.~\cite{hoyos_protecting_2011}. First, it was found that any tiny breaking of the perfect coupling-field correlation drives the system to IRFP physics.
However, when the perfect correlation in the Hamiltonian Eq.~\eqref{eq:Hcorrelated} is maintained, weak disorder is {\it{irrelevant}} for the clean critical point, and quite large disorder is required to drive the system towards a non-trivial line of critical points, where unusual properties emerge, such as an increase of the entanglement entropy with the disorder strength. These numerical results from Ref.~\cite{hoyos_protecting_2011} are reproduced in Fig.~\ref{fig:correlated} (a). 

Model Eq.~\eqref{eq:Hcorrelated} is an interesting example where by construction the disordered system is always strictly critical at the local level, satisfying the condition $J_i=h_i$ in Eq.~\eqref{eq:TFI}, and thus naturally yielding $\delta=0$. This apparent suppression of local randomness protects the clean physics against small disorder $D$, but at strong enough $D$ a new physics appears where entanglement increases with $D$.
The effective central charge, extracted from the logarithmic growth in the main panel of Fig.~\ref{fig:correlated} (a), is shown in the inset. Note that an extension to an interacting XXZ  version was studied in Ref.~\cite{getelina_entanglement_2016}, reaching similar conclusions as compared to the above non-interacting situation.

\paragraph*{(ii) Rainbow and Randbow states---}
There is another family of engineered disordered models which has motivated an important number of works: the so-called rainbow model~\cite{vitagliano_volume-law_2010,ramirez_conformal_2014}, and its extensions, in particular the "Randbow" XX chain~\cite{alba_unusual_2019}
\be
{\cal{H}}_{\rm Randbow~XX}=\sum_{i=-L}^{L+1}J_i\,{\rm{e}}^{-h|i|}\left(S_i^xS_{i+1}^x+S_i^yS_{i+1}^y\right).
\label{eq:randbow}
\ee
For the disorder-free ($J_i=\rm constant$) case, the spatial structure of its inhomogeneity, exponentially decaying from the center of the chain, allows to
apply the SDRG rule and construct the ground-state: the concentric singlet phase depicted in Fig.~\ref{fig:correlated} (c). 
Analyzing its entanglement
properties~\cite{ramirez_conformal_2014,ramirez_entanglement_2015,rodriguez-laguna_more_2017}, a volume-law scaling emerges with the entropy proportional to the number of sites inside the subsystem. This remains true for any non-zero value of the exponentially decaying parameter $h$, with the particularly interesting volume-law asymptotic scaling~\cite{rodriguez-laguna_more_2017}, in the limit $h\ll 1$ and $h\ell\gg 1$
\be
S_{\rm vN}(h,\ell)\approx \frac{1}{6}\ln\left(\frac{{\rm{e}}^{\ell h}-1}{h}\right)\sim \frac{h}{6}\ell.
\ee

The introduction of a true randomness in the couplings $J_i$ (on top of this exponentially decaying pattern) has led Alba {\it{et al.}}~\cite{alba_unusual_2019} to the so-called Randbow case, with the following results for the asymptotic forms, at large $\ell$
\be
\label{eq:F2D}
S_{\rm vN}(\ell)\propto \left\{
\begin{array}{lr}
\frac{\ln 2}{6}\ln \ell \quad&{\rm{if}}~h=0~\rm (RSP)\\
\ell \ln 2\quad&{\rm{if}}~h\to \infty~\rm (Rainbow)\\
\sqrt{\ell}\quad&{\rm{otherwise}}~\rm (Randbow).
\end{array}
\right.
\ee
It is remarkable to observe that the RSP scaling only survives in the limit $h=0$ of no decaying couplings. In the opposite limit, the rainbow concentric singlet phase can only overcome the effect of disorder in $J_i$ for a "vertically decaying" inhomogeneity $h\to \infty$. Finally, the entire regime $0<h<\infty$ falls in the intermediate situation, the so-called "Randbow" phase, see Fig.~\ref{fig:correlated} (d), with an unusual $\sqrt\ell$ area-law violation. This exotic scaling is a direct consequence of the ground-state structure: exponentially rare "rainbow" regions having long-distance singlets, coexist with "bubble" regions (made of short-range singlets) having a power-law decaying probability~\cite{alba_unusual_2019}.

Let us finally comment on the effect of interactions in the XXZ version of the randbow chain. While irrelevant for the RSP physics, here there very structure of the SDRG iterations lead to the fact that the above area-law violation appears to be specific to the free-fermion point. From SDRG calculation, attraction is found to restore the volume-law scaling, while repulsive interactions induce a strict area-law scaling~\cite{alba_unusual_2019}.

\section{Many-body localization probed by quantum entanglement}
\label{sec:mbl}

\subsection{Area {\it{vs.}} volume law entanglement for high-energy eigenstates}
Entanglement is a key concept to gain some insight on many-body localization (MBL) physics, briefly described in Section~\ref{subsec:mbl}, see also Refs.~\cite{nandkishore_many-body_2015,abanin_recent_2017,alet_many-body_2018,abanin_many-body_2019} for recent reviews.
In isolated quantum systems, thermalization implies that the system acts as its own heat bath. This is the case for the so-called ergodic regime, adjacent of the MBL phase, see Fig.~\ref{fig:gs_phase_diag} (c) where the eigenstate thermalization hypothesis (ETH)~\cite{deutsch_quantum_1991,srednicki_chaos_1994} is expected to hold.  In this delocalized phase, the reduced density matrix  of a high-energy eigenstate can be interpreted as an equilibrium (high-temperature) thermal density matrix. Therefore, the entanglement entropy of such a highly excited eigenstate must be very close to the thermodynamic entropy of the subsystem at high temperature, thus exhibiting a volume-law scaling. Such delocalized infinite-temperature eigenstates are usually well described by random states having a maximal entanglement entropy~\cite{page_average_1993}. 

\begin{figure}[h!]
\begin{center}
\includegraphics[width=.85\columnwidth,clip]{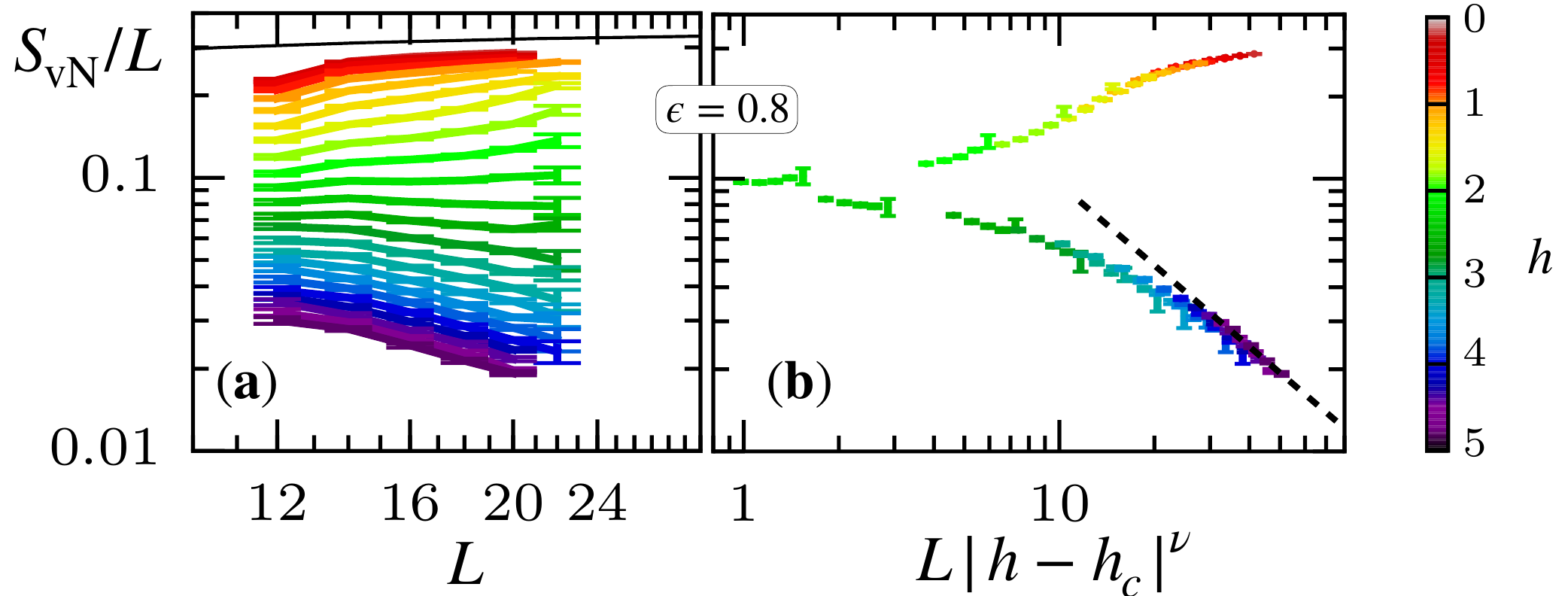}
\caption{Entanglement entropy density $S_{\rm vN}/L$ for the MBL problem at high energy. Shift-invert exact diagonalization results for half-chain cuts performed over periodic Heisenberg chains with a random field Eq.~\eqref{eq:H}, obtained for high-energy ($\epsilon=(E-E_{\rm min})/(E_{\rm max}-E_{\rm min})=0.8$) eigenstates with various chain lengths $L=12,\ldots, 22$. In panel (a) a clear qualitative change is visible upon increasing disorder $h$, from volume-law (black line shows the Page's law~\cite{page_average_1993}) to area-law, with a critical point observed for $h_c\sim 2.5$. Panel (b) shows a scaling plot obtained with $h_c=2.27$ and $\nu=1$. The dashed line $\sim 1/L$ represents the strict area-law situation. Figure adapted from Luitz {\it{et al.}}~\cite{luitz_many-body_2015}.}
\label{fig:volumearea}
\end{center}
\end{figure}

Volume-law entanglement at high temperature has been clearly observed for clean quantum spin chains~\cite{sorensen_quantum_2007,sato_computation_2011,alba_eigenstate_2015,keating_spectra_2015,vidmar_entanglement_2017}, as well as in the ergodic side of weakly disordered chains~\cite{bauer_area_2013,kjall_many-body_2014,luitz_many-body_2015,luitz_ergodic_2017}. In contrast, the MBL regime violates ETH and eigenstates display a much weaker area-law entanglement, quantitatively closer to the entanglement entropy of a ground-state~\cite{eisert_colloquium:_2010,dupont_many-body_2019}. 
Such qualitatively distinct properties have been observed numerically in various studies~\cite{kjall_many-body_2014,luitz_many-body_2015,lim_many-body_2016,khemani_critical_2017}. In order to illustrate this, Fig.~\ref{fig:volumearea} shows exact diagonalization results for the half-chain von-Neumann entanglement entropy, obtained together with D. Luitz and F. Alet in Ref.~\cite{luitz_many-body_2015} for the random-field Heisenberg chain model Eq.~\eqref{eq:H}. When $S_{\rm vN}$ is normalized by the system size, the transition from volume- to area-law is clearly visible around $h_c\sim 2.5$ (random fields are drawn from a box $[-h,h]$) at this energy density $\epsilon=0.8$, see also the scaling plot in panel (b). Our numerical data are compatible with a volume-law entanglement  at criticality~\cite{grover_certain_2014}, and with a strict area-law scaling in the MBL regime, shown as a dashed line in Fig.~\ref{fig:volumearea} (b). 
Note that in the MBL phase, Bauer and Nayak~\cite{bauer_area_2013} reported a weak logarithmic violation of the area law for the maximum entropy, obtained from the (sample-dependent) optimal cut, see also~\cite{kennes_entanglement_2015,dupont_many-body_2019}.\\

\subsection{Distributions of entanglement entropies}

\subsubsection{Distribution across the ETH-MBL transition}

\begin{figure}[t!]
\begin{center}
\includegraphics[width=\columnwidth,clip]{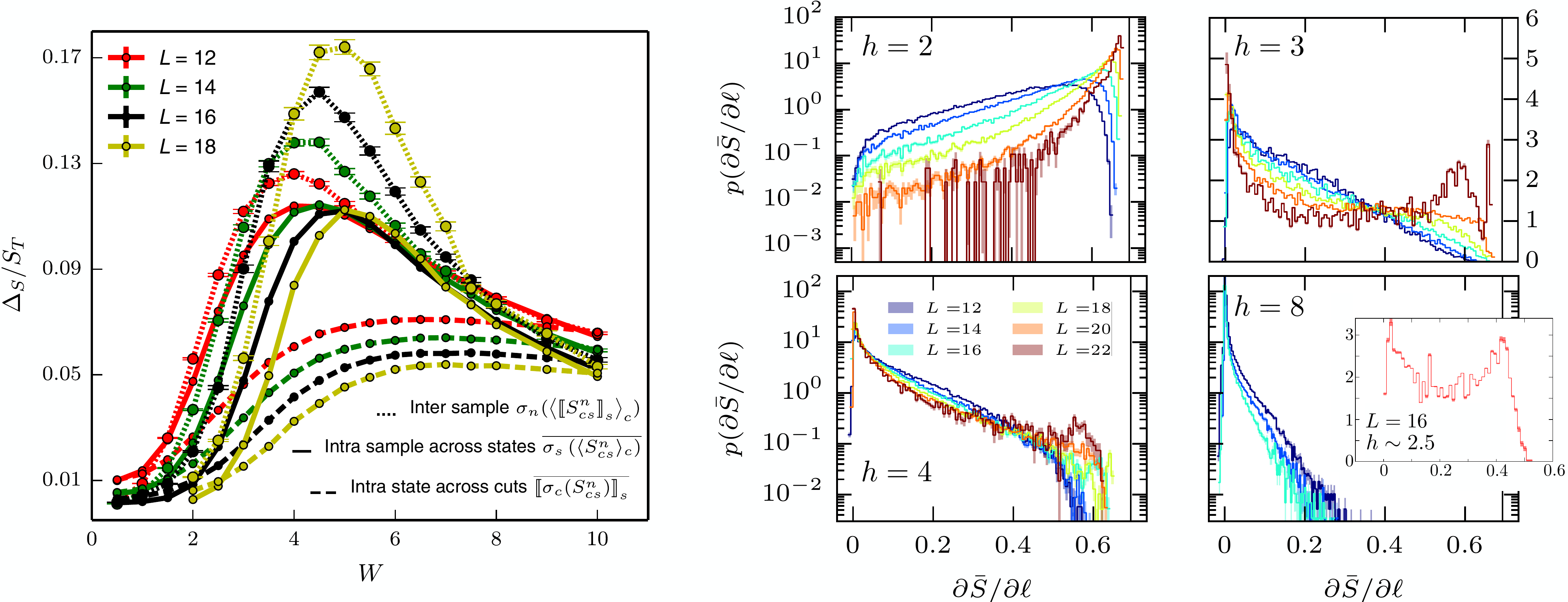}
\caption{Entanglement entropy distributions across the ETH-MBL transition. Left: from Khemani {\it{et al.}}~\cite{khemani_critical_2017},  (normalized) standard deviation of the von-Neumann entropy $\Delta_S/S_T$ plotted against  disorder strength $W$ for the standard model Eq.~\eqref{eq:H} with additional second neighbor exchange (yielding a critical disorder strength $W_c\sim 7$).  In the critical region, $\Delta_S$ is dominated by sample-to-sample fluctuations. Figure taken  from Ref.~\cite{khemani_critical_2017}.  Right: Distribution of entanglement slopes, from Yu {\it{et al.}}~\cite{yu_bimodal_2016}, for model Eq.~\eqref{eq:H}. 
Upon increasing the disorder strength $h$, there is a clear qualitative change in the distributions. When the transition is approached, a bimodal shape is identified, a structure also observed at the level of a single disordered sample (inset) for 6000 eigenstates. Figure taken from Ref.~\cite{yu_bimodal_2016}.}
\label{fig:S_variance}
\end{center}
\end{figure}

In order to go beyond the disorder and eigenstate average entropies, a systematic study of their distributions turns out to be extremely instructive, as first discussed in Refs.~\cite{bauer_area_2013,kjall_many-body_2014,luitz_many-body_2015,lim_nature_2016}. An enhancement of the variance with increasing system sizes $L$ was reported when approaching the critical region, thus providing a quantitative tool, see for instance Fig.~\ref{fig:S_variance} (left).
Another very thorough and exhaustive study was provided by Yu {\it{et al.}}~\cite{yu_bimodal_2016} for the standard-model Eq.~\eqref{eq:H}, see Fig.~\ref{fig:S_variance} (right) where the four panels show a remarkable qualitative change in the distributions of entanglement slopes upon increasing the disorder.  In addition, a bimodal structure was found at criticality, a feature surprisingly observed also for a  {\it single} disorder realization (see inset, where the distribution is computed from eigenstates in the {\it{same}} disorder sample). As argued by Khemani {\it{et al.}}~\cite{khemani_critical_2017,khemani_two_2017}, a key for understanding the MBL transition may come from the differences between fluctuations of entanglement coming from different eigenstates in the {\it same} disordered sample, as compared to fluctuations coming from different samples, see Fig.~\ref{fig:S_variance} (right) taken from Ref.~\cite{khemani_critical_2017}.

\subsubsection{Strong disorder distributions}

\begin{figure}[h!]
\begin{center}
\includegraphics[width=\columnwidth,clip]{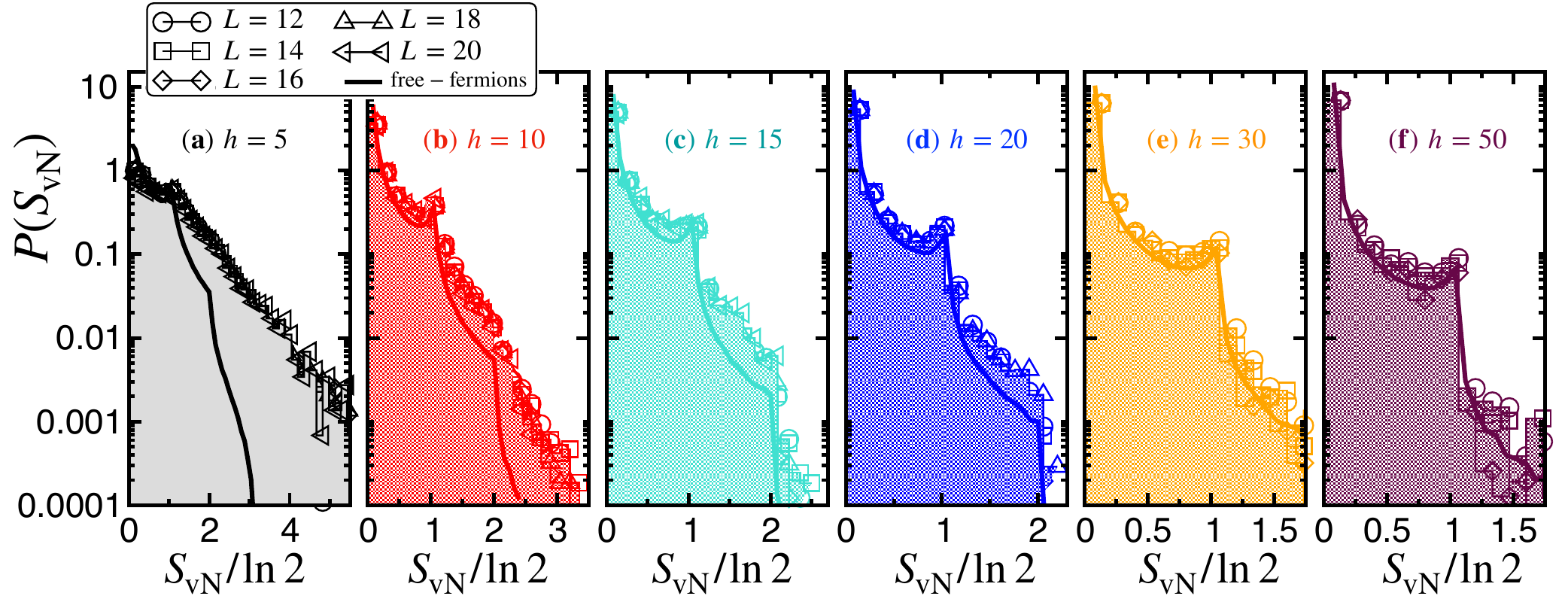}
\caption{Strong disorder behavior of the half-chain entanglement entropy distributions for the random-field Heisenberg chain model  Eq.~\eqref{eq:H}, deep in the MBL regime. Shift-invert ED results for highly excited states at ${{\epsilon}}=0.5$.
Different panels (a-f) show histograms of $S_{\rm vN}/\ln 2$  collected for $L=12,\,14,\,16,\,18,\,20$ (different symbols) over several thousands of independent random samples for varying disorder strengths $h=5,\,10,\,15,\,20,\,30,\,50$, as indicated on the plot. The non-interacting (free-fermions) case for $L=32$ is also shown (lines) for comparison. One sees the peaked structure gradually developing when $h$ increases. Note the quasi-absence of finite-size effects.}
\label{fig:histo_mbl}
\end{center}
\end{figure}

At strong disorder, deep in the MBL regime the entanglement entropy is obviously very small. However, following our previous discussion for the non-interacting case (Section~\ref{sec:SD} and Fig.~\ref{fig:ps_anderson}), it is also instructive to take a look at the histograms in the interacting case at large disorder. Fig.~\ref{fig:histo_mbl} displays several panels for $P(S_{\rm vN})$ at various disorder strengths $h=5,\,10,\,15,\,20,\,30,\,50$, computed for $L=12,\,14,\,16,\,18,\,20$ at infinite temparature $\epsilon=0.5$. One can observe the following remarkable effects: 
\begin{enumerate}
\item[(i)]{Finite size effects are almost absent, confirming the fact that the localization length is very small deep in the MBL phase~\cite{mace_multifractal_2019,laflorencie_chain_2020}.} 
\item[(ii)]{Upon increasing $h$, the influence of interactions becomes gradually less visible, clearly noticeable when comparing the MBL data (symbols) with the non-interacting case (full lines, data from panel (b) of Fig.~\ref{fig:ps_anderson}). A qualitative difference is only apparent below $h\approx 10$, when more pronounced at $h=5$ when the MBL-ETH transition is approached.}
\item[(iii)]{The peaked structure is also clearly present, signalling anomalously weakly disordered sites. We have also checked that the probability $\rho_1=P(|S_{\rm vN}/\ln 2-1|\le 0.05)$ decays $\sim h^{-1}$, like in the non-interacting case. One can therefore anticipate that the entanglement entropy will be dominated by such "rare" events.}

\end{enumerate}

\newpage\section{Concluding remarks}
\label{sec:conclusion}
In this Chapter, the entanglement properties of various disordered quantum chains have been discussed, with a global focus on the von-Neumann entanglement entropy $S_{\rm vN}$ for three different classes of random spin chains. Extensive numerical results have been presented, and reviewed together with an important  literature on this topic.

For Anderson localized XX chains in a random magnetic field, $S_{\rm vN}$ exhibits universal scaling, with different forms which depends on the energy. Nevertheless, it was shown that there is a unique length scale which controls the real space localization of single particle states and the scaling functions of the many-body entanglement entropy. For very strong randomness, the behavior of the distributions is also remarkable, showing some peculiar features which clearly capture some salient low and high energy properties.

A second set of systems that we discussed concerns infinite randomness physics. For random-bond XX chains at zero temperature, we unveiled a nice finite-size crossover for the effective central charge, controlling the logarithmic scaling of the von-Neumann entropy, from the clean behavior to the random-singlet asymptotic form. As another example of infinite randomness, the quantum Ising chain was studied at and away from criticality, for both zero and infinite temperature. The logarithmic critical scaling is similar (and therefore universal) at all energies, with only a non-universal constant which depends on the energy.

We have also reviewed on the existing results beyond free fermions, {\it{e.g.}} random singlet phases with higher spins, and also discuss the cases of engineered disordered systems with locally correlated randomness or the so-called rainbow/randbow chain models.

Finally the strongly debated problem of many-body localization has also been discussed through the properties displayed by eigenstates entanglement entropies at high energy. Going beyond the volume-law to area-law paradigm for the ETH-MBL transition, the shape of the distributions have been investigated and discussed for all regimes, including strong disorder where Anderson and MBL insulator displays almost similar entanglement structure, despite their clearly different dynamical response~\cite{znidaric_many-body_2008,bardarson_unbounded_2012,serbyn_universal_2013,vosk_many-body_2013,andraschko_purification_2014}.

\section*{Acknowledgments}
It is a great pleasure to thank all my collaborators on this topic of entanglement properties in quantum disordered systems:
Fabien Alet, Maxime Dupont, Jos\'e Hoyos, Gabriel Lemari\'e, David Luitz, Nicolas Mac\'e, Eduardo Miranda, Andr\'e Vieira, Thomas Vojta. 
This work was supported by the Agence Nationale de la Recherche under programs GLADYS ANR-19-CE30-0013, and ANR-11-IDEX-0002-02, reference ANR-10-LABX-0037-NEXT.

%

\end{document}